\documentstyle[prl,aps,epsf]{revtex} 
\begin{document} 
\renewcommand{\textfraction}{0.6}
\renewcommand{\dbltopfraction}{0.4}
\renewcommand{\bottomfraction}{0.4}
\renewcommand{\dblfloatpagefraction}{0.4}
\topmargin= -0.1 true in
\draft 
 
\def\bfs{{\bf s}}   \def\bfp{{\bf p}}
\def\calS{{$\cal S$}} \def\calP{{$\cal P$}} \def\calSp{{$\cal S'$}}
\def\calL{{$\cal L$}} 
\def\nmax{{n_{max}}} \def\nsb{{n_{10}}}

\twocolumn[\hsize\textwidth\columnwidth\hsize\csname %
@twocolumnfalse\endcsname
\phantom{xx}
\vskip -1cm
\hfill{{\it Preprint} NCU/CCS-1998-1010; NCHC-phys-1998-1024; NSC-CTS-981001}

\medskip

\title
{Mean-Field HP Model, Designability and Alpha-Helices in Protein Structures}

\author{C.T. Shih$^1$, Z.Y. Su$^1$, J.F. Gwan$^1$, 
B.L. Hao$^{2,3}$, C.H. Hsieh$^1$ and H.C. Lee$^{3,4}$} 
\address{
$^{1}$National Center for High-Performance Computing, Hsinchu, Taiwan, ROC\\
$^{2}$Inst. of Theoretical Physics, Academia Sinica, Beijing, China\\
$^{3}$National Center for Theoretical Sciences, Hsinchu, Taiwan, ROC\\
$^{4}$Dept. of Physics and Center for Complex Systems, 
National Central University, Chungli, Taiwan, ROC}

\date{Received December 14, 1998; revision received \today}

\maketitle
\begin{abstract}
Analysis of the geometric properties of a mean-field HP model on a
square lattice for protein structure shows that structures with large
number of switch backs between surface and core sites are chosen
favorably by peptides as unique ground states.  Global comparison of
model (binary) peptide sequences with concatenated (binary) protein
sequences listed in the Protein Data Bank and the 
Dali Domain Dictionary indicates that the highest
correlation occurs between model peptides choosing the favored
structures and those portions of protein sequences containing
alpha-helices.

\end{abstract}
]

\medskip
 
The three-dimensional structure of proteins is a complex physical and
mathematical problem of prime importance in molecular biology,
medicine and pharmacology \cite{cre92}.  It is believed that the
folding instruction of a protein is encoded in its amino acid sequence
\cite{anfinsen73} and from model studies much has been learned about
protein structure and folding kinetics \cite{dill,mani,li96,shak98}.
Yet much still remains to be understood.  This simple fact is already
intriguing: the number of possible globular structures for a peptide
of typical length - about 300 amino acids - is practically infinite;
the number of proteins whose structures are known empirically or
hypothetically is more than a hundred thousand and is growing rapidly
with time; the number of classes of native protein structures is about
five hundred and is believed unlikely to exceed a thousand in the long
run \cite{cre92,DDD}.  Numerical simulations based on lattice models
have shown that structures of exceptionally high {\it designability} -
those that attract a large number of protein sequences to conform to
it - do exist \cite{li96,shak98,design}.  Why such structures would
emerge is however not well understood.  Protein folding also has an
outstanding temporal feature: the initial collapse to globular shape
and the formation of $\alpha$-helices are completed in less than
$10^{-7}$ seconds \cite{Munoz}, while the rest of the folding takes up
to ten seconds to complete.

In this report, based on results from a mean-field lattice model we
observe that structures with high designability are preponderant in a
type of substructure that suggests $\alpha$-helices in real proteins
and we explain the reason for this phenomenon.  This notion is
supported by global comparisons of model structural sequences with
(binary) sequences constructed from sets of proteins of known
structure: the Protein Data Bank (PDB) \cite{PDB} and the Dali Domain
Dictionary (DDD) \cite{DDD}.  Since the mean-field in the model
represents the hydrophobic potential that is known to cause the
initial collapse of a peptide to a globular shape, the results may
explain why the initial collapse {\it and} the formation of
$\alpha$-helices occur essentially simultaneously and rapidly, and are
temporally separated from other slower folding processes that are
driven by far-neighbor inter-residual interactions.

In the minimal model for protein folding, the HP model of 
Dill {\it et al.} \cite{dill}, the 20 kinds of amino acids are divided
into two types, hydrophobic and polar.  This reduces a peptide chain
of length $N$ to a binary ``peptide'' ${\bf p} = (p_1,p_2,\ldots,p_N)$,
where $p_i= $0 (1) if the amino acid at the $i$th position on the
chain is polar (hydrophobic).  The structure of a protein is
represented by a self-avoiding path compactly embedded on a lattice
\calL, and the energy associated with a peptide conforming to a
particular structure is computed from the contact energies between the
nearest-neighbor residues that are not adjacent along the peptide.  A
set of well tested contact energies derived from proteins of known
structure is the Miyazawa-Jernigan matrix, which is however well
approximated by an effective mean-field potential expressing the
hydrophobicities of the residues \cite{miyazawa96}.  In the binary
form of this approximation the Hamiltonian of the HP model is 
reduced to that of a mean-field model:
\begin{equation}
H(\bfp, \bfs)  =  -{\bf p}\cdot{\bf s} =  
\frac12 (|\bfs -\bfp|^2 - {\bf p}^2 - {\bf s}^2) 
\label{e:hp2}
\end{equation}
where ${\bf s}= (s_1,s_2,\ldots,s_N)$ is a binary ``structure''
converted from a self-avoiding path with the assignment: $s_i= 1$ (0)
if the $ith$ site is a core (surface) site on the lattice
\cite{li96,li98}.  Empirical observation suggests that protein folding
proceeds in two steps, a first stage of fast collapse and formation of
alpha-helices (and probably some not properly folded beta-sheets) 
presumably caused mainly by hydrophobic interactions under 
polymeric constraints, followed by a second stage of slow annealing
caused by far-neighbor inter-residue interactions that gives the final
native state \cite{Munoz,Williams}.  Since Eq.(\ref{e:hp2}) is a
local, mean-field approximation that leaves out residual - 
{\it i.e.}, left over from mean-field averaging - far-neighbor
interactions, it can be relied on to account for only the first stage.

We denote by \calS\ the set of all distinct structures $\bfs$ on
\calL\ and by \calP\ the set of all possible peptides \bfp\ of length
$N$.  For each \bfp\ the selection of the \bfs\ giving the minimum $H$
defines a mapping from \calP\ to \calS.  There are \bfp's that are
mapped to more than one \bfs's or are mapped to \bfs's that correspond
to more than one self-avoiding paths.  Such \bfp's are removed from
\calP\ and their target \bfs's are removed from the competition for
high designability, because a peptide that does not conform to a
unique structure at all times is not expected to survive the
evolutionary selection process \cite{unique}.  It has also been shown
that not admitting degenerate states in a coarse-grained model is
similar to removing peptides that has low foldability in a
finer-grained model \cite{buch99}.  (Many states that are degenerate
in the present coarse-grained model would in a model with higher
energy resolution be states of different symmetries with nearly
degenerate energies.  The energy landscape for the ground state among
these would likely contain deep local minima and a peptide choosing
such a ground state would likely be a poor folder.)  The mapping then
partitions what remains in \calP\ into classes, with all the \bfp's in
each class mapped to a single \bfs, whose designability is simply the
number of \bfp's in the class.  For this mapping the right-hand-side
of Eq.(\ref{e:hp2}) reduces to being proportional to $|\bfs-\bfp|^2$,
the Hamming distance the two points \bfp\ and \bfs\ in an
$N$-dimensional unit hypercube.  Then the designability of an \bfs\ is
equal to the Voronoi polytope around it in this hypercube \cite{li98}.

Excepting those removed for degeneracy, \calP\ is just the set of all
the vertices - on a unit hypercube.  In comparison, owing to the
constraints of compactness and self-avoidance imposed on paths on
\calL, points in \calS\ are sparsely distributed in the hypercube so
that ${\cal S}\subset{\cal P}$.  For example, on a $6\times 6$ square
lattice, the number of elements in (including those to be removed for
degeneracies) \calP\ is $2^{36}= 68719476736$, while those in \calS\
is 30408 (but only 18213 of them have no path degeneracy).  
If the points in \calS\ were uniformly distributed in the
hypercube, then the Voronoi polytope around each \bfs\ would be the
same and every \bfs\ would have the same designability.
But owing to boundary effects and geometric constraints
imposed on the compact paths on \calL, the distribution of \bfs's in
\calS\ {\it cannot} be uniform, those \bfs's residing in regions in
the hypercube that are of especially low density (in \bfs's) will then
have especially high designability.

We now examine how geometric constraints cause the emergence of \bfs's
with especially high designabilities by first attempting to replace
the constraints by a set of explicit algebraic ``rules''.  Consider
a structure in \calS\ to be a chain of 0's and 1's linked by $N-1$
links of three types, 0-0, 1-0 or 0-1, 1-1, with $n_{00}$, $n_{10}$ and
$n_{11}$ being the numbers of such links, respectively.  The structure
is partitioned by the 1-0 links into $n_{10}+1$ ``islands'' of
contiguous 1's or 0's.  (Peptides in \calP\ may be similarly
described, but the only constraint on any \bfp\ is that the total
number of 0's and 1's be $N$.)  For \calL\ being a square lattice with
side $L$, two of the most important constraining rules are: ($i$) A
single 0 may only occur at an end of a path; ($ii$) An isolated single
1 may only either occur at or be one 0-island away from an end of a
path.  Space does not allow us to give more than another relatively
simple example (with $L>4$): For a path having the pattern $\bfs=
(1\cdots1)$ (both the ends of the path are 1-sites), $2n_{00}+n_{10}=
8L-8$ and $2\le n_{10} \le 4L-12$.  It is in fact extremely difficult
if not impossible to exhaust the complete set of such rules needed to
reduce \calP\ to \calS.  For our purpose it suffices to identify a
large enough set of rules which reduces \calP\ to a ${\cal S'}$ that
is a sufficiently close to ${\cal S}$ for us to understand the
origin and characteristics of structures of high designability.

\begin{figure}[m]
\epsfxsize= 8.5cm\epsfbox{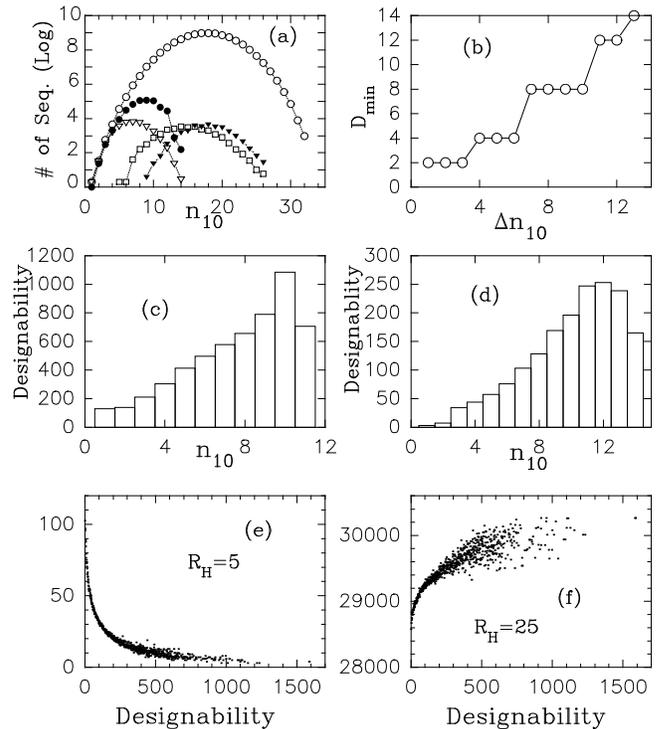}
\caption{(a) Number of peptides vs. $n_{10}$ for \calP\ (open circle),
\calP$_4$ (solid triangle), and \calP$_5$ (square), 
and number of structures vs. $n_{10}$ for \calSp\ (solid circle)
and \calS\ (open triangle),  on the $6\times 6$ lattice. 
(b) Smallest Hamming distance vs. differences of $n_{10}$
of all the 30408 paths in \calP\ on the $6\times 6$ lattice.
Average designabilities of the paths vs.  $n_{10}$ 
for the (c) $4\times 7$ and (d) $6\times 6$ 
lattices, respectively.  Number of neighboring structures 
within a Hamming distance of (e) $R_H$=5 and (f) $R_H$=25 of 
a structure of given designability.}
\label{f:constraint}
\end{figure}

In Fig.\ref{f:constraint}(a) the number of elements in \calP (open
circle), \calSp\ (solid circle) and \calS\ (open triangle) on a
$6\times 6$ lattice are respectively plotted against $n_{10}$. The
total number of elements under the curve for \calP\ gives the total
number of sites in the hypercube.  \calSp\ is slightly greater than
\calS\ but is much smaller than \calP.  (The boundary of \calSp\ owes
its roughness to the incompleteness of the set of rules used to
construct it.)  It is seen that whereas for \calP\ the maximum 
possible value
for $n_{10}$ is $4L -4 = 32$, for \calS\ and \calSp\ the corresponding
maximum is much less: $\nmax =14$.  As $n_{10}$ approaches $\nmax$
from below, the number of elements in \calSp\ decreases rapidly
whereas those in \calP\ increases toward a maximum.  It happens that
in the hypercube the smallest Hamming distance between two structures is
approximately proportional to the difference in their respective
$n_{10}$ numbers.  This is evident in Fig.\ref{f:constraint}(b), where
the smallest Hamming distance is plotted against the difference in
$n_{10}$ for all the pairs among the 30408 binary structures
on a $6\times 6$ lattice, and is consistent with results given in
\cite{onuchic98} in which $x(p)$ (the degree of clustering of
hydrophobic residues) is analogous to $n_{10}$.

Since allowed structures with $n_{10}$'s having values close to
$\nmax$ live in a region of the hypercube that is also most heavily
populated by peptides, it follows that they would on average
have a large Voronoi polytope, and hence are most likely to have the
highest designabilities.  This is substantially borne out by the
results shown in Figs.\ref{f:constraint}(c) and (d) computed for the
allowed structures on the $4\times 7$ and $6\times 6$ lattices,
respectively, where average designability is plotted against 
$n_{10}$.   The average designability does not exactly peak at
$n_{10}=\nmax$ but rather at $n_{10}$'s just less than $\nmax$.  Why
this should be so is not yet clearly understood.  Structures with maximum
$n_{10}$ are the most constrained and are very few in number so that 
otherwise secondary details might have had a larger effect on their
designabilities.  Preference for large $n_{10}$'s has also been
observed in other 2D and 3D lattices.   The relation between 
high designability and sparse population is further illustrated in 
Figs.\ref{f:constraint}(e) and (f), where the number of structures 
within a Hamming distance $R_H$ of a given structure is plotted against 
the designability of that structure.   In (e), where $R_H$=5, it is 
seen that structures with high designability have far fewer near 
neighbors than structures with low designability.  In (f), 
where $R_H$=25, it is seen that all structures have approximately 
the same large numbers of near and far neighbors. 

Now something interesting emerges.  A structure with its $n_{10}$
(almost) maximized but not allowed to have single 1's or 0's except at
its ends (rules ($i$) and ($ii$)) will have a preponderance of the
4-mer (1100) in the interior, so that large stretches of it will have
the form $(\cdots11001100\cdots)$ which suggests the linear structure
of $\alpha$-helices on a lattice.  A corollary is that structures
with core to surface ratios close to unity are favored by
designability.  This implies a diameter of approximately 10 residues
for an ideal protein, which is consistent with the typical size of 300
to 1000 amino acids in natural proteins.  Note that the selection of
structures with maximized $n_{10}$ is a consequence of the geometric
property of the Hamiltonian (\ref{e:hp2}) in hyperspace and does not
depend on the specifics of a lattice.  That larger $n_{10}$'s are
favored is a notion qualitatively consistent with the conclusion 
drawn from recent studies on folding kinetics that optimal 
structures are also minimally frustrated \cite{frust}.

To see if what we have observed so far has anything to do with real
proteins we compare five sequences, \calP$_{1-5}$, 
each being a concatenation of a set of real protein or 
($6\times 6$) lattice binary
peptides: ${\cal P}_1$, (a) the representative
non-redundant 2886 proteins (sequence similarity smaller than 90\%)
\cite{sander98} culled from the 9257 entries in PDB \cite{PDB}, 
or (b) the even less redundant set of 1394 entries of protein domains 
from DDD \cite{DDD}, converted to binary sequences
based on hydrophobicity \cite{radzicka88}; ${\cal P}_2$, the sections
in ${\cal P}_1$ that fold into $\alpha$-helices; ${\cal P}_3$, the
sections in ${\cal P}_1$ that fold into $\beta$-sheets; ${\cal P}_4$,
the 27006 peptides in \calP\ mapped to the 15 structures of the
highest designabilities; ${\cal P}_5$, the 24134 peptides in \calP\
mapped to the 1545 structures of the lowest designabilities.
Interestingly the H/P ratios of the five sequences are all very close
to 1; the percentage of hydrophobic residues contained in each is
respectively $50.00\%$, $49.75\%$, $56.18\%$, $50.43\%$ and $49.05\%$
for ${\cal P}_1$ (from PDB) through ${\cal P}_5$.  Fig.\ref{f:constraint}(a)
shows that neither distribution of peptides in \calP$_4$ (solid 
triangle) and \calP$_5$ (open square) vs. $n_{10}$ is random, which
corroborates the results of \cite{stat}.  In particular, in \calP$_5$
(\calP$_4$) peptides with larger (smaller) $n_{10}$'s are slightly 
favored over those with smaller (larger) $n_{10}$'s.

\begin{figure}[m]
\epsfxsize= 8.5cm\epsfbox{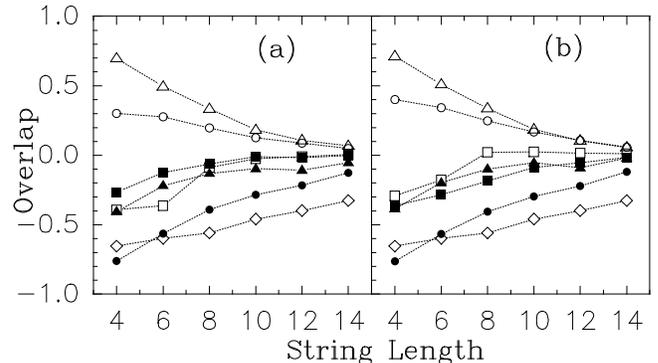}
\caption{Overlap of 
frequency distribution functions of lattice peptides 
and PDB proteins (a) and DDD protein domains (b) as a function of 
word length $l$: 
$O^{(l)}_{14}$ (open circle), $O^{(l)}_{15}$ (solid circle), 
$O^{(l)}_{24}$ (open triangle), $O^{(l)}_{25}$ (solid triangle), 
$O^{(l)}_{34}$ (open square), $O^{(l)}_{35}$ (solid square) and
 $O^{(l)}_{45}$ (open diamond).}  
\label{f:overlap}
\end{figure}

Let $f^{(l)}_i(m)$ be the frequency, normalized to that of 
a sequence with an H/P
ratio of unity (if the word has $n_H$ H's and the actual frequency
of the word is $f$, then the normalized frequency is 
$(n_H/n_P)^{-n_H}f$), of the $m$th binary word
of length $l$ occurring in sequence
${\cal P}_i$ and let $F^{(l)}_i(m) = (f^{(l)}_i(m)- {\bar
f}^{(l)}_i)/Z$ be the normalized frequency distribution function,
where ${\bar f}^{(l)}_i= 2^{-l}\sum_{m}f^{(l)}_i(m)$ is the mean
frequency and $Z = (\sum_{m}(f^{(l)}_i(m)- {\bar f}^{(l)}_i)^2)^{1/2}$
is the norm.  The relations $\sum_{m}F^{(l)}_i(m)= 0$ and
$\sum_{m}(F^{(l)}_i(m))^2= 1$ hold.  The pairwise overlaps
$O^{(l)}_{ij}= \sum_{m=1}^{2^l} F^{(l)}_i(m)F^{(l)}_j(m);\
i=1,2,3;\ j=4,5$  
that measure correlations between ${\cal P}_i$ and ${\cal P}_j$ for
$l$=4-14 are given in Figs.\ref{f:overlap}(a) and (b), where the real protein 
sequences used are from PDB and DDD, respectively.   
The two sets of overlaps are qualitatively similar.  
It is seen that ${\cal
P}_4$ (${\cal P}_5$) is positively (negatively) correlated with ${\cal
P}_1$ and ${\cal P}_2$.  For all values of $l$ the strongest
correlation occurs between the model sequence of high designability
(${\cal P}_4$) and the real protein sequence rich in $\alpha$-helices
(${\cal P}_2$).  The sequence of high designability is poorly
correlated with the sequence rich in $\beta$-sheets (${\cal P}_3$)
and, as expected, the strongest anti-correlation occurs between the
two model sequences of high (${\cal P}_4$) and low (${\cal P}_5$)
designabilities.   

Even though the favoring of surface-core repeats by peptides folding
into high-designability structures is most likely not lattice specific
(provided the H/P ratio is close to one), the particular choice of the
(1100) repeat has the characteristic of a square lattice.  For
instance, on a hexagonal lattice the predominant repeat would more
likely be (10) rather than (1100).  There is some justification for
selecting a square lattices over a hexagonal because in real proteins
the backbone does not favor small-angle bends.  On the other
hand, real proteins do not live on lattices and the equivalent of (10)
repeats does occur in real proteins where $\beta$-sheets are exposed
to solvent.  Thus the low correlation between ${\cal P}_3$ and ${\cal
P}_4$ is to some extent an artifact of the square lattice, and it may
be better to interpret the (1100) repeats on a square lattice as
representing $\alpha$ type {\it and} some $\beta$ type repeats 
(but not the latter's foldings) in real proteins.  

Our study suggests that the rough formation of $\alpha$-helices and
some $\beta$-sheets and the collapse of proteins into globular shapes
are primarily determined by hydrophobicity.  Since only the mean-field
part of the inter-residue interaction is included in the model, this
implies that details of the residual inter-residue interaction that
determine the final shape of the native state are not important at
this stage.  It has been pointed out that structures of high
designability in a lattice model with two-letter amino acid alphabet 
may not be especially designable for higher-letter
alphabets \cite{buch99}.  Although the situation may be different on a
lattice larger than the $5\times 5$ lattice used in \cite{buch99}, it
does remain to be verified whether our findings persist in
finer-grained and more realistic models.  If it does, then we can 
better understand why the formation of $\alpha$-helices and the
collapse would happen on a similar time scale, of the order
$10^{-7}$s, why the formation of $\beta$-sheets would take somewhat
longer (about $10^{-6}$s) \cite{Munoz,Williams}, and why these time
scales would be so much shorter than the time needed to complete the
rest of the folding ($10^{-1}$s to $10$s).  This scenario is in any 
case consistent with the finding in a recent statistical analysis of
experimental data: local contacts play the key role in
fast processes during folding \cite{plax98}.

\medskip
This work is partly supported by grants NCHC88-CP-A001 to 
ZYS from the National Center for High-Performance Computing, and 
NSC87-M-2112-008-002 to HCL and NSC87-M-2112-007-004 to BLH from the 
National Science Council (ROC).  HCL thanks Simon Fraser University for
hospitality in the Summer of 1998 during which part of the paper was
written.


\end{document}